\documentclass[pra,aps,superscriptaddress,twocolumn,showpacs,nofootinbib]{revtex4-1}

\usepackage{placeins}
\usepackage{amsfonts, amsmath,amssymb,amsthm,bm}
\usepackage{mathrsfs}
\usepackage{graphicx}
\usepackage[usenames,dvipsnames]{color}
\usepackage{palatino}
\usepackage{fancyhdr}
\usepackage{mathrsfs}
\usepackage{multirow}
\usepackage{hyperref}
\usepackage{bigints}
\usepackage{xcolor}

\usepackage[framemethod=TikZ]{mdframed}

\hypersetup{
   colorlinks=true,
   linkcolor=blue,
   citecolor=blue,
   urlcolor=Violet
}

\definecolor{grau}{rgb}{0.95, 0.95, 0.95}

\newmdenv[%
    backgroundcolor=grau,
    linecolor=black,
    outerlinewidth=1pt,
    roundcorner=2mm,
    skipabove=\baselineskip,
    skipbelow=\baselineskip,
    innertopmargin=5pt,
    innerbottommargin=5pt
]{bozzed}

\newcommand{\N}{\mathbb{N}}

\theoremstyle{plain}
\newtheorem{theorem}{Theorem}
\newtheorem{lemma}[theorem]{Lemma}
\newtheorem{observation}[theorem]{Observation}
\newtheorem{definition}[theorem]{Definition}
\theoremstyle{definition}

\begin{document}
\title{Free agency and determinism: is there a sensible definition of computational sourcehood?}
\author{Marius Krumm}
\affiliation{Institute for Theoretical Physics, University of Innsbruck, Technikerstra\ss e 21a, A-6020 Innsbruck, Austria}
\author{Markus P.\ M\"uller}
\affiliation{Institute for Quantum Optics and Quantum Information, Austrian Academy of Sciences, Boltzmanngasse 3, A-1090 Vienna, Austria}
\affiliation{Vienna Center for Quantum Science and Technology (VCQ), Faculty of Physics, University of Vienna, Vienna, Austria}
\affiliation{Perimeter Institute for Theoretical Physics, 31 Caroline Street North, Waterloo, ON N2L 2Y5, Canada}

\begin{abstract}
Can free agency be compatible with determinism? Compatibilists argue that the answer is yes, and it has been suggested that the computer science principle of ``computational irreducibility'' sheds light on this compatibility. It implies that there cannot in general be shortcuts to predict the behavior of agents, explaining why deterministic agents often appear to act freely. In this paper, we introduce a variant of computational irreducibility that intends to capture more accurately aspects of actual (as opposed to apparent) free agency: computational sourcehood, i.e.\ the phenomenon that the successful prediction of a process' behavior must typically involve an almost-exact representation of the relevant features of that process, regardless of the time it takes to arrive at the prediction. We argue that this can be understood as saying that the process itself is the source of its actions, and we conjecture that many computational processes have this property. The main contribution of this paper is technical: we analyze whether and how a sensible formal definition of computational sourcehood is possible. While we do not answer the question completely, we show how it is related to finding a particular simulation preorder on Turing machines, we uncover concrete stumbling blocks towards constructing such a definition, and demonstrate that structure-preserving (as opposed to merely simple or efficient) functions between levels of simulation play a crucial role.
\end{abstract}

\date{May 30, 2023}

\maketitle

\section{Introduction}
\label{SecIntroduction}
Do humans, animals, some machines, or other systems or processes have some sort of control over their actions that deserves to be called ``free will''? This question has been discussed in various forms in the philosophical literature over the last two millenia (see e.g.\ Ref.~\cite{SEPFreeWill} for an introduction). At first sight, the fact that agents are subject to the same laws of nature as stones and atoms seems to be in tension with an intuitive understanding of free will: after all, when an agent has taken a decision, they will fundamentally never have had the ``freedom to do otherwise''. Moreover, one might be inclined to think that it is not the agent, but rather the laws of nature (and initial conditions) that represents the ``source'' of the agent's actions. While quantum theory suggests that the fundamental laws of nature are best understood as probabilistic rather than deterministic, it has been convincingly argued~\cite{DennettElbowRoom,Dennett,Pinker} that decisions are not free simply because they are random. Therefore, the question of compatibility of determinism and free will remains relevant even in the face of indeterminism (for the complementary question of compatibility of free agency and \emph{indeterminism}, see e.g.~\cite{MuellerBriegel}).

Among the various positions in the debate, compatibilism~\cite{FischerRavizza,Dennett} amounts to the claim of a positive answer to this question.
Compatibilists have argued against the incompatibility of free will and determinism in a variety of ways: by rejecting the idea that the freedom to do otherwise is necessary for free will; by claiming that the freedom to do otherwise is compatible with determinism; or by formulating ways in which sourcehood can be ascribed to an agent even in a deterministic world (for an overview, see Ref.~\cite{SEPCompatibilism}).

In a complementary development, there have been proposals to analyze the relation between determinism and free will via theoretical computer science. After all, we do not live in an \emph{arbitrary} deterministic (or probabilistic) world, but in a world that seems to comply with physical versions of the Church-Turing thesis~\cite{Gandy,Arrighi}. This suggests to treat decision-making systems (including agents) and predictors as \emph{algorithms} in some sense, and to contemplate aspects of free will with information-theoretic notions and methods.

Two such approaches (described in more detail in Section~\ref{SecNoShortcut} below) offer an explanation for why some physical systems (in particular human agents) \emph{appear} to have free will. One approach, computational irreducibility~\cite{Wolfram,Zwirn1,Zwirn2,Zenil}, demonstrates that the behavior of many physical systems cannot be predicted without simulating every single step of their time evolution in full detail. Another approach, based on time complexity arguments~\cite{Lloyd}, shows that it takes typically more time to predict an agent's actions than it takes the agent to come to its decision by itself. In a nutshell, \emph{decision making systems do not in general admit shortcuts}. However, as already noted by Bringsjord~\cite{Bringsjord}, this insight in itself may explain some of the \emph{phenomenology} of free will, but it does not have much to say about whether agents are \emph{actually} free.

In this paper, we suggest to study a variant of computational irreducibility that is intended to formalize an aspect of \emph{actual} free will more directly: a computational notion of \emph{sourcehood}. In a nutshell, we ascribe computational sourcehood to a computable process $P$ if attempts to reproduce its outputs can typically only succeed by running an almost-exact step-by-step simulation that contains all functionally relevant aspects of $P$. In contrast to computational irreducibility, we do not focus on the time it takes to perform the simulation, i.e.\ computational sourcehood claims that such ``cloning'' of the simulated process is also necessary if the simulation takes an arbitrarily long time. Thus, if we regard the collection of all relevant abstract information-theoretic elements of $P$ as (part of) an agent, then it is always \emph{this agent} that is invoked when $P$'s behavior is generated, reproduced or predicted. We conjecture that this kind of sourcehood is indeed ``typical'', i.e.\ that it can be attributed to a large variety of computational processes.

We will follow a careful terminology choice also made by M\"uller and Briegel~\cite{MuellerBriegel}: we will mostly avoid talking about ``free will'' (unless when we follow other authors), and use the notion of ``free agency'' instead, in order to avoid arousing associations with specifically human aspects of this notion. As M\"uller and Briegel put it, \emph{``In philosophy, free will is mostly tied to specifically human traits, such as being the proper subject of moral praise and blame, or a capacity for conscious deliberation or for the linguistic expression of self-reflective thought.''} These traits are irrelevant for the purpose of this paper, since we will concentrate on a single specific notion: that of \emph{sourcehood}, and the question of whether a technical, formal definition of it can be found. It is clear that the specifically human notions mentioned above can play no role in such a definition.

For similar reasons, details of the human decision-making process (as described, for example, by neuroscience or biology) will play no role in our analysis, and so will questions of semantics as discussed in general theories of information~\cite{Burgin}. Similarly as Lloyd's, our analysis will operate on a more abstract level, and the in-principle applicability of our results to physical systems, including humans, can be motivated by the simple observation that \emph{``$[\ldots]$ because the known laws of physics can be simulated on a computer, the dynamics of the brain can be simulated by a computer in principle --- it is not necessary that we know how to simulate the operation of the brain in practice.''}~\cite{Lloyd}. A different motivation for our approach (and those by Lloyd and Wolfram) may come from versions of the computational theory of mind~\cite{CompTheoryOfMind}. The purpose of our paper is not to argue for any of those views, but to consider a specific technical question about algorithms that can be motivated by them.

Our article is organized as follows. In Section~\ref{SecNoShortcut}, we give a brief summary of previous ``no-shortcut'' approaches to apparent free agency: computational irreducibility and time complexity arguments. We introduce the idea of computational sourcehood and its difference to computational irreducibility with a thought experiment (``John the cook'') in Section~\ref{SecJohn} and more formally in Section~\ref{SecCompSourcehood}. Section~\ref{SecRigorous} contains the technical results of this paper: successively improved attempts at formally defining a version of computational sourcehood that is non-trivial, meaningful, and has the chance to lead to a sensible formulation of our main conjecture (that many processes are the computational sources of their behaviors). Finally, we conclude in Section~\ref{SecConclusions}.

\section{No-shortcut approaches to apparent free agency}
\label{SecNoShortcut}
In this section, we briefly summarize two approaches intended to explain the phenomenology of free agency using a computational perspective. Both approaches have in common that they identify the huge effort necessary to predict an agent's decisions as an account of apparent freedom. As we will see, the approaches differ in how they expose the difficulty of agent predictions. While the first approach (which is closely related to ours) focuses on the non-existence of shortcuts and simplifications in the prediction process, the second approach focuses on computation time and uncomputability.

\subsection{Computational irreducibility}
One concept in computer science that has been suggested to shed new light on the relation between free agency and determinism is \emph{computational irreducibility}, proposed by S.\ Wolfram~\cite{Wolfram}.

To explain this concept, let us consider the success of scientific predictions. Most physical systems are extremely complex objects constructed from a vast number of smaller parts. Nonetheless,  often the behavior of crucial properties of such complex systems can be described with a few equations. For example, in mechanics, many of the considered physical systems are extended objects. Such systems consist of countless atoms, which themselves are built from smaller particles. However, as we know from our mechanics lectures, predicting the evolution of the mechanical properties of these systems often does not require a simulation of all the individual parts. E.g.\ in astronomy, the orbits of gas planets can be well approximated via Kepler's laws, without the need to consider all the gas molecules involved in the movement of the planet.

These considerations teach us that the behavior of some important properties of complex physical systems can be predicted without having to model all their microscopic features. This implies that these predictions are possible because the simulation allows for massive shortcuts and simplifications. Now, Wolfram's concept of \emph{computational irreducibility} refers to the observation that not all physical properties and complex systems allow for such shortcuts and simplifications. In other words, there exist physical questions that require a near-perfect simulation of all the details involved to be answered. While Wolfram formulates this notion for computable processes, he follows a strong intuition that our physical world does indeed correspond to a computation of some sort. Physical versions of the Church-Turing thesis aim at substantiating such intuitions in different possible ways~\cite{Gandy,Arrighi}.

In his book~\cite{Wolfram}, Wolfram suggests the possibility that agents, including humans, show behavior that is computationally irreducible, even if the agent is deterministic. In other words, answering some questions about the future of an agent might require full simulation of all functionally relevant details of the agent's thought process and environment. Wolfram argues that this might be the origin of apparent free will:

``\emph{And it is this, I believe, that is the ultimate origin of the apparent freedom of human will. For even though all the components of our brains presumably follow definite laws, I strongly suspect that their overall behavior corresponds to an irreducible computation whose outcome can never in effect be found by reasonable laws.}''

This explanation of free will was critized by S. Bringsjord~\cite{Bringsjord} as being ``epistemologically correct'', but ``metaphysically wrong'', in particular:

``\emph{If someone's will is \emph{apparently} free, it hardly follows that that will is \emph{in fact} free. Nowhere in ANKS \emph{[his book]} does Wolfram even intimate that he maintains that our decisions are in fact free.}''

We will reconsider this distinction, and the role of computational notions in its analysis, in Section~\ref{SecCompSourcehood}, where we take it as an inspiration to introduce a modified version of computational irreducibility.
 
While Wolfram discusses the phenomenology and implications of computational irreducibility and provides examples in terms of cellular automata, he does not give an exact formalization of this notion in Ref.~\cite{Wolfram}. Such a formal definition has been proposed via Turing machines by H.\ Zwirn and J.-P.\ Delahaye~\cite{Zwirn1,Zwirn2}. In a nutshell, they call a function $f:\N\to\N$ computationally irreducible if all efficient Turing machines computing $f(n)$ will essentially also have to compute $f(1),\ldots,f(n-1)$. While this corresponds to a straightforward implementation of the main properties of Wolfram's examples (in particular if $f(n)$ encodes the $n$th row of a cellular automaton), it is not clear whether this formulation is the most suitable one for the study of free agency. And since many choices of detail have to be made in the construction of the definition, it is not clear how many functions actually satisfy it.

\subsection{Lloyd's time complexity argument}
Secondly, we consider S.\ Lloyd's~\cite{Lloyd} idealization of agents as Turing machines, more specifically as computable deciders. Such deciders are Turing machines that map a description of a decision problem to a yes/no-answer, or fail to come to a conclusion. More formally, a (computable) decider $d$ is identified with a Turing machine that receives an input string $k$, and outputs $d(k) \in \{0,1\}$ or fails to halt, i.e. $d(k)$ undefined. Lloyd considers a function $f(d,k)$ that is supposed to predict the answers of all such deciders $d$ for all inputs $k$. By adapting the proof of the halting problem, Lloyd argues that such a function must be uncomputable:

``\emph{The unpredictability of the decision-making process does not arise from any lack of determinism --- the Turing machines involved could be deterministic, or could possess a probabilistic guessing module, or could be quantum mechanical. In all cases, the unpredictability arises because of uncomputability}.''

Nonetheless, as Lloyd points out, decisions in real environments usually have to be made within a limited amount of time. Therefore, he considers time limited deciders and argues that for such time limited deciders a program predicting all decisions is computable. However:

``\emph{In summary, applying the Hartmanis-Stearns diagonalization procedure shows that any general method for answering the question `Does decider $d$ make a decision in time $T$, and what is that decision?'\ must for some decisions take strictly longer than $T$ to come up with an answer. That is, any general method for determining $d$'s decision must sometimes take longer than it takes $d$ actually to make the decision.}''

Using the diagonalization argument to discuss self-reference of universal deciders, Lloyd arrives at the following conclusion about free will:

``\emph{Now we see why most people regard themselves as possessing free will. Even if the world and their decision-making process is completely mechanistic --- even deterministic --- no decider can know in general what her decision will be without going through a process at least as involved as the decider's own decision-making process. In particular, the decider herself cannot know beforehand what her decision will be without effectively simulating the entire decision-making process. But simulating the decision-making process takes at least as much effort as the decision-making process itself.}''

\section{John the cook: a thought experiment}
\label{SecJohn}
To set the stage, consider the following thought experiment. Suppose that John Smith is a very talented (and emotional) cook. Every morning, he decides what kind of breakfast to prepare, but his repertoire of meals is very large. On some mornings, he might remember his late Canadian wife and then prepare a particular omelette with cranberries and maple syrup. On other mornings, he might prepare a Pho soup in memory of his travels to Vietnam. The number of different breakfasts he might cook extends into the thousands.

Suppose we would like to predict what John is going to eat tomorrow morning. Furthermore, suppose that our prediction task is made particularly easy by assuming that physics is perfectly deterministic, and, in fact, computable and discrete at some microscopic scale. Since we know that John is going to spend the evening and night in his apartment, we are going to build a huge machine around his house: it scans the apartment to perfect accuracy, and then simulates his apartment (including John's brain and body) on an extremely powerful computer. We would like our simulation to tell us John's choice of breakfast well before dawn: it is certainly more fun to make a prediction before the predicted event has happened.

\begin{figure}[hbt]
\begin{center}
\includegraphics[width=.5\textwidth]{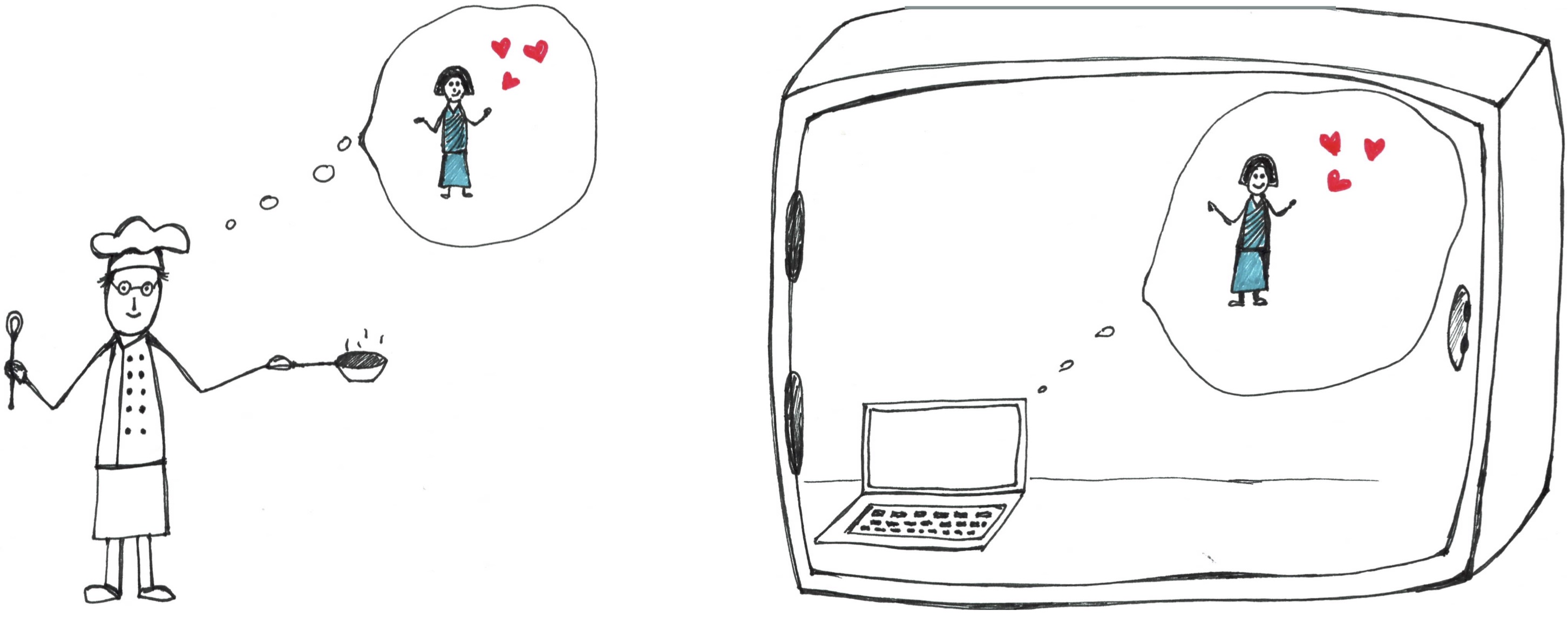}
\caption{A universal computer (in the safe, on the right) reproduces the outputs of another process, i.e.\ its observable actions (John preparing breakfast, on the left). Computational sourcehood means that this prediction cannot typically be successfully performed without representing all relevant elements (here: thoughts, emotions) of that process in the simulation.}
\label{fig_john}
\end{center}
\end{figure}

However, we are familiar with the phenomenon of computational irreducibility and are thus warned that the simulation may have to be \emph{really detailed} in order to succeed. Furthermore, Lloyd's arguments~\cite{Lloyd} apply, so we should be prepared that our simulation finishes only \emph{after} John has arrived at his decision. Therefore, we take precaution: directly after scanning John and his apartment, before starting the simulation, we put the scan data and the computer into an immensely secure safe.

If the simulation finishes before breakfast, we can directly confront John with our successful prediction. Otherwise, we can invite John to our laboratory and let him witness how we open the safe and obtain the delayed prediction. In both cases, we might then be inclined to say (unjustifiably, as we argue below): \emph{``See, John? You think that \emph{you and your emotions} were irreducibly involved in the decision to prepare the Canadian omelette, but what happened in the safe was only determined by your (and your apartment's) physical state yesterday night. Your thoughts and emotions this morning had no impact on the decision whatsoever!''}

But would this denial of free agency indeed by justified? We argue that the answer to this question does \emph{not} depend on any possible time delay in this thought experiment. The crux of the situation --- that reproducing the decision seems in principle possible without an element that John regards as a crucial aspect of his free will --- is the same \emph{regardless of the simulation time}. But this suggests that questions of efficiency should ultimately be irrelevant for the problem of free agency.

We argue that the relevant question is a different one. An important aspect of John's impression to make a free, autonomous choice is that the particular class of emotions and thoughts in his mind are in some meaningful sense the \emph{source} of the decision, the choice of breakfast. That is, not only have they been involved in the causal chain of physical events leading to a particular decision, but their \emph{presence seems indispensable} to ultimately arrive at the decision. This motivates us to ask a specific question about the simulation in the safe: \emph{did the simulation necessarily contain correlates of these emotions and thoughts?} In other words, did the simulation --- and does \emph{every} successful simulation --- also have to ``think about John's late wife'' in some sense?

If the answer is ``no'' and such simulations can typically predict the decision by \emph{completely unrelated means} --- say, either by a drastic shortcut or by a completely different method --- then John may rightly be worried. On the other hand, a positive answer to this question opens up the possibility to ascribe John sourcehood for his decisions. As is standard for some compatibilist positions~\cite{SEPCompatibilism}, we can do so by interpreting these thoughts and emotions as parts of John's identity. In other words: if we identify \emph{John the agent} not with the particular matter that encompasses his brain in the apartment, but with the \emph{collection of functionally relevant structures represented in his brain}, then \emph{this agent is the source of its decisions}: the simulated process in the safe has simply contained another representation of John.

It is not the purpose of this paper to make these conceptual ideas and arguments philosophically fully rigorous. Instead, the above merely serves as a \emph{motivation} to consider a specific technical question in computer science: can we find a formal definition of a notion of ``computational sourcehood'' which expresses the property above in a sensible way? That is, if an algorithm simulates the behavior of another one, does it typically have to reproduce its features exactly? We argue that the possibility of formalization of this idea is a necessary condition for even considering any philosophically advanced and detailed instantiation of these ideas: the impossibility of formalization is often an important indicator that the motivating ideas cannot hold water.

\section{Computational sourcehood}
\label{SecCompSourcehood}

Let us begin by rephrasing some aspects of the ``John the cook'' thought experiments in computer science terminology. Think of John as a Turing machine (TM) $T$, with the configuration of John's apartment and the content of his brain yesterday night as the input, and think of the computer in the safe as a universal TM $U$. (Here we focus on TMs only because it is the most well-known model of computation, and we think that our results are independent of the choice of model that is used to implement computations.) John's observable actions (in particular, his choice of breakfast) are to be found among $T$'s outputs, and $U$ is able to reproduce these outputs exactly. Does this tell us anything about \emph{the way how} $U$ arrives at these outputs?

Intuitively, universal TMs simulate other TMs $T$ exactly and step by step; as we will explain below, this is certainly true for all universal TMs that are described in standard textbooks. Now, if this is true for the computer in the safe, then the answer to the above question is positive: \emph{yes, the computer has reproduced copies of the neural correlates (representations) of John's thoughts and emotions.} But is this so by necessity?

Universal TMs $U$ can be fed a description of another TM $T$ together with some input $x$, and they can use this to compute the output of $T$ on input $x$. It seems extremely hard to imagine that there would be any other way for $U$ to do so rather than by simulating $T$ step by step --- with perhaps some rare exceptions (some TMs $T$ may have obvious inefficiencies that can easily be shortcut by $U$). Thus, it is natural to conjecture that there is a large class of Turing machines $\mathbf{T}$ such that all universal TMs have to resort to some form of step-by-step simulation to simulate them on most inputs:
\begin{bozzed}
\textbf{Informal Conjecture.} There is a large class of Turing machines $\mathbf{T}$ that are ``the source of their own actions'' in the following sense. Consider any TM $U$ that is \emph{universal}, i.e.\ that can emulate every other TM $T$: that is,
\begin{equation}
   U(p_T,x)=T(x)
   \label{eqSim}
\end{equation}
for every TM $T$ and every input $x$ on which $T$ halts, where $T\mapsto p_T$ is an effective description of the TM $T$. Then, for all $T\in\mathbf{T}$, the universal TM $U$ will generate the output $U(p_T,x)$ for \emph{most} inputs $x$ in such a way that it contains during its simulation, in suitable form and at suitable time steps, exact copies of the subsequent states that the machine $T$ takes during its computation on input $x$.

We may then speak of either $T$, or a given pair $(T,x)$, as an instance of ``computational sourcehood''.
\end{bozzed}
Establishing a formal version of this conjecture would allow us to reason that John can be viewed in a specific sense as the ``source of his own actions'', assuming that the Turing machine $T$ he implements is contained in the set $\mathbf{T}$. But regardless of the problem of free agency, establishing or disproving formal versions of the above might yield interesting insights into the nature of universal computation.

Clearly, this notion of computational sourcehood is closely related to computational irreducibility: both concepts claim that an exact representation is necessary for prediction. However, computational irreducibility focuses more on the impossibility to speed up a computation, whereas computational sourcehood claims more generally that every successful prediction of the results of a computation must involve an exact image of the original, \emph{even slow ones}. This implies in particular that no essential speedups are possible, but it makes additional claims about the structure of simulations, including those that take substantially longer time.

\section{Towards a rigorous formulation of the conjecture}
\label{SecRigorous}
This section assumes familiarity with some basic notions of computer science like Turing machines, prefix codes, and computability; for an introduction, see e.g.~\cite{Hennie,Hopcroft,AroraBarak,Papadimitriou}. To fix some notation, we will denote the set of natural numbers by $\mathbb{N}=\{0,1,2,3,\ldots\}$ (which includes zero), and the set of finite binary strings by
\[
   \{0,1\}^*=\{\varepsilon,0,1,00,01,10,11,000,\ldots\},
\]
where $\varepsilon$ is the empty string. The length of a string $s\in\{0,1\}^*$ will be denoted $\ell(s)$, with $\ell(\varepsilon)=0$. In general, if $\Sigma$ is any finite alphabet, the set of finite words over $\Sigma$ will be denoted $\Sigma^*$.

\subsection{Textbook universal Turing machines behave as conjectured}
\label{SubsecTextbook}
It is a straightforward yet cumbersome exercise to formally construct a universal TM in a rigorous way. We will now sketch a construction given in the textbook by Hennie~\cite{Hennie}; for another choice, see e.g.\ Hopcroft et al.~\cite{Hopcroft}.

TMs come in different versions. What all definitions have in common is a finite set $Q$ of internal states of the machine, including an identified \emph{initial state} $q_0$ (and sometimes a \emph{final state} $q_f\neq q_0$). Every TM has a finite number $n$ of \emph{tapes}, such that each tape has a set of \emph{cells} that is either indexed by the natural numbers $\mathbb{N}=\{0,1,2,\ldots\}$ (a ``one-way infinite'' tape) or by the integers $\mathbb{Z}=\{\ldots,-2,-1,0,1,2,\ldots\}$ (a ``two-way infinite'' tape). Every tape cell contains one symbol of a finite alphabet $\Sigma$, and there is a special ``blank symbol'' $\#\in\Sigma$ that is carried by all but finitely many tape cells. For every tape, the TM has a \emph{tape head} that moves along that tape and points at a particular cell at any given time.

Single steps of operation are determined by a (perhaps partial) \emph{transition function}, which we may write as
\[
   \delta:(q,\vec\sigma)\mapsto (q',\vec\sigma',\vec d).
\]
We interpret this as follows. At every time step, if the TM is in internal state $q$ and reads the symbols $\vec\sigma=(\sigma_1,\ldots,\sigma_n)$ on its $n$ tapes (the content of the cells where the tape heads are pointing), then it replaces the tape contents by $\vec\sigma'=(\sigma'_1,\ldots,\sigma'_n)$, transitions into the new internal state $q'$, and moves the head on tape $i$ either to the left (if $d_i=L$), to the right (if $d_i=R$), or not at all (if $d_i=N$).

The initial state of the TM -- in particular, the way that the input is supplied to the tapes -- depends on the convention of choice, except that the initial internal state is always assumed to be $q_0$. For our purpose, we assume that all but a finite number of tape cells must initially carry the blank symbol $\#$, so that the input is finite in this sense. If, at any time step, the TM reaches the state $q_f$, then it ``halts'', and the content of one or all of the tapes is -- according to some choice of convention -- interpreted as the TM's \emph{output}. Depending on the choice of convention, there may also be other events that are interpreted as ``halting'' (in particular, if the TM is not assumed to have a distinguished state $q_f\in Q$): for example, trying to turn left from cell $0$ of a one-way infinite tape, or having $\delta$ undefined on the current combination of $q$ and $\vec\sigma$.

Hennie~\cite{Hennie} makes a particular choice of convention for defining TMs and for constructing a universal TM. He considers TMs that have a single ($n=1$) one-way infinite tape, \emph{no} final state $q_f$, and a finite alphabet $\Sigma=\{0,1,\ldots,k\}$, where $0$ denotes the blank symbol. The set of internal states is denoted $Q=\{0,1,\ldots,m\}$, where $q_0:=0$ is the initial state. The tape head direction can be either left ($L=0$) or right ($R=1$), but the head cannot stay at its current position. The TM is assumed to halt if it either runs off the tape (to the left of cell $0$), or if it lands in a combination of internal state $q$ and tape symbol $\sigma$ such that $\delta(q,\sigma)$ is undefined. Hennie is not particularly explicit in defining what the ``input'' or ``output'' of the TM computation shall be (some function of the initial resp.\ final tape pattern): the choice of convention for how to do so can be adapted to the desired context.

Values of the transition function like
\[
   \delta(q,\sigma)=(q',\sigma',d)
\]
are represented via quintuples
\[
   (q,\sigma,\sigma',d,q').
\]
Hence, every TM $T$ can be represented as a finite list of such quintuples. We will consider this list to be ordered, and by convention to start with a quintuple that begins with the starting state $q_0$. We now obtain the first step of Hennie's construction of a universal TM: the choice of description $p_T$ of a TM $T$. The description $p_T$ will be a binary string, namely a unary representation of the quintuples that define $T$. This is best explained by one of Hennie's examples. Consider the TM described by the quintuples
\begin{eqnarray*}
(q_0, 1, 0, R, g_1),&&\enspace (q_1,0,1,R,q_2),\enspace (q_1, 1,1,R,q_1),\\
(q_2, 0,0,L, q_3), &&\enspace (q_3,0,0,R,q_0), \enspace (q_3, 1,1,L,q_3).
\end{eqnarray*}
Recalling that internal states and directions are also integers, we get the equivalent (but less readable) form
\begin{eqnarray*}
(0,1,0,1,1),&&\enspace (1,0,1,1,2),\enspace (1,1,1,1,1),\\
(2,0,0,0,3),&&\enspace (3,0,0,1,0),\enspace (3,1,1,0,3).
\end{eqnarray*}
Denoting $k$ consecutive $1$'s by $1^k$, the unary representation of a quintuple is
\[
   (a,b,c,d,e)\mapsto 1^{a+1}0 1^{b+1} 0 1^{c+1} 0 1^{d+1} 0 1^{e+1},
\]
and we will describe a sequence of quintuples by concatenating the descriptions of each quintuple, separated by pairs of blank symbols (zeros). That is, our representation of the example TM becomes
\begin{eqnarray*}
1011010110110011010110110111001101101101101100\\
111010101011110011110101011010011110110110101111.
\end{eqnarray*}
In addition to the description $p_T$ of the TM $T$, we also have to supply our universal TM $U$ with the tape content $x$ of TM $T$. Formally,
\[
   U(p_T,x)=U(\langle p_T,x\rangle),
\]
where $\langle p_T,x\rangle$ is the initial tape content of $U$ that by definition encodes the pair $p_T$ and $x$. Note that we cannot simply define $\langle p_T,x\rangle$ to be the concatenation of $p_T$ and $x$: first, the machine needs to know where $p_T$ ends and $x$ starts; and second, the alphabet $\Sigma_U$ of $U$ will in general be different from the alphabet $\Sigma_T$ of $T$ (potentially containing fewer elements), so that $x$ cannot in general be directly copied from $T$'s tape to that of $U$.

Hennie chooses the following construction. Given the initial tape content $x=x_1 x_2 x_3 \ldots$ of $T$ (with $x_i\in\Sigma_T$), consider the unary encoding
\[
   x_u:=1^{x_1+1} 0 1^{x_2+1} 0 1^{x_3+1}\ldots
\]
By construction, the TM $U$ shall contain three distinguished ``marking symbols'' $A,B,C\in\Sigma_U\setminus\{0,1\}$. Then, the full encoding is defined as
\[
   \langle p_T,x\rangle=A 0^{|Q_T|+|\Sigma_T|+2} B p_T 000C x_u.
\]
In summary, the cells between $A$ and $B$ represent a \emph{buffer region} that is large enough to contain a unary representation of $T$'s current internal state $q$ and currently scanned tape symbol $\sigma$. The region between $B$ and $C$ contains the description of $T$, and the rest of the tape contains a description of $T$'s initial tape content.

In essence, the TM $U$ is constructed to work as follows. It copies the current description of $T$'s internal state $q$ (which is what succeeds the $B$ marker) and the current description of $T$'s scanned tape cell $\sigma$ (succeeding the $C$ marker) into the buffer region. Then $U$ scans the $p_T$ region to find the quintuple beginning with the combination $(q,\sigma)$ inscribed in the buffer region (if no such quintuple is found then $U$ halts) and moves the $B$ marker in front of that quintuple. Finally, $U$ reads the direction $d\in\{L,R\}$ and new tape symbol description $\sigma'\in\Sigma_T$, updates $T$'s tape description accordingly, and moves the $C$ marker either left or right (if it moves left then it checks whether this would imply that simulated $T$ runs off the tape; if so $U$ halts). Then it restarts this cycle.

For a more detailed description of $U$, see \cite[Sec.\ 2.3]{Hennie}. In principle, to define $U$, one would have to give the detailed definition of its transition function $\delta_U$ (or, equivalently, its long list of defining quintuples). This would be extremely cumbersome. Hence, what is done instead is to argue that subroutines like ``copying'' or ``searching'' can be incorporated into the definition of $U$ by effectively using other TMs as ``submachines''. This is explained in~\cite[Sec.\ 2.1]{Hennie}.

More generally, in the rest of the paper, we will often simply describe in words how a TM is supposed to work on the contents of its tapes, and assume that this can in fact be implemented in the definition of the TM in some way. In addition to the possibility of checking the validity formally for each single case, such reasoning is standardly justified via the Church-Turing thesis.

To formulate in (semi-formal) detail how Hennie's construction satisfies our conjecture, let us introduce one additional piece of notation. Given any TM $T$, input $x$, and time $t\in\mathbb{N}$, denote by
\[
   \mathcal{C}_T(x,t)
\]
the configuration of $T$ after having computed for $t$ steps on input $x$. With ``configuration'', we mean a complete description of the content of its tapes, of its internal state $q$, and of the location of its tape head(s).
The (countable) set of all possible configurations of TM $T$ will be denoted $\mathcal{C}_T$. The countable set of all configurations that any Turing machine of this sort can hold is denoted $\mathcal{C}:=\bigcup_{T\,\,\mathrm{ TM}}\mathcal{C}_T$; this is the set of all configurations where the control is in some integer state, the tape head points somewhere, and finitely many tape cells are filled by integers (and the rest with blanks). This allows us to formulate the following observation.
\begin{bozzed}
\begin{observation}
\label{ObsHennie}
Hennie's universal TM $U$ has the following property. There exists a ``simple'' function $\varphi:\mathcal{C}_U\to\mathcal{C}$ and, for every TM $T$, an increasing function $\tau_T:\mathbb{N}\to\mathbb{N}$ such that
\[
   \mathcal{C}_T(x,t)=\varphi\left(\strut\mathcal{C}_U(\langle p_T,x\rangle,\tau_T(t))\right).
\]
In other words, the time evolution of $T$ on input $x$ can ``easily be read off'' from the time evolution of $U$ on inputs $p_T$ and $x$, with a possible slowdown $t\mapsto \tau_T(t)$.

Moreover, the function $\tau_T$ has a simple characterization in the following way. If we follow the step-by-step evolution of $U(p_T,x)$ for times $\tau=1,2,3,\ldots$, then we can directly observe whether $\tau=\tau_T(t)$ for some $t$ or not. Namely, if $U$ has returned to the beginning of its simulation cycle at time $\tau$, then this is the case, and otherwise not. We can determine $t$ by starting at $t=0$, and by increasing $t$ by one whenever $U$'s computation restarts its cycle.
\end{observation}
\end{bozzed}
In other words, Hennie's universal TM works exactly as described in our Informal Conjecture: it simulates every TM $T$ step by step, and reproduces exact images of the computational state of $T$ at suitable time steps.

In most textbooks, emphasis is placed on the fact that the simulation of $T$ by $U$ leads to a slowdown that is at most polynomial; in our notation, this means that the function $t\mapsto \tau_t$ grows at most polynomially. However, for our purpose, this fact is not particularly relevant: as discussed earlier, we are not interested in the time it takes to arrive at a prediction, but in the information that is involved in generating the prediction.

The function $\varphi$ is essentially an extended ``snapshot function'' of the simulated TM~\cite{AroraBarak}. It can easily be described in words: the obtain $T$'s internal state $q$ at time $t$, just look at the buffer region to the right of symbol $A$ at time $\tau_t$. To see the position of the tape head, simply search for the marker $C$ and count groups of ones; to determine the content of type cell $i$, count $i$ groups of ones right of the left-most $000$ and translate the corresponding $1^{x_i+1}$ into $x_i$. This is a very ``simple'' function; yet, in what sense a general definition of ``simple'' should apply to \emph{all} universal computers is a non-trivial question that we will address next.

\subsection{A simulation preorder on TMs}
Do \emph{all} universal TMs satisfy a version of Observation~\ref{ObsHennie}, as Hennie's universal TM does? To address this question, we first need to give a general definition of a universal TM $U$. In Section~\ref{SecCompSourcehood}, we have intuitively thought of universal TMs as those that satisfy Eq.~(\ref{eqSim}). Note however that Hennie's universal TM does \emph{not} satisfy that equation, but rather
\begin{equation}
   T(x)=\psi\left(U(p_T,x)\right),
   \label{eqNotSim}
\end{equation}
where $\psi$ is another ``simple function'' that extracts $T$'s final tape content from that of $U$. Namely, $U$ does not exactly end with the same output as $T$, but with a certain unary \emph{encoding} of its output, preceding by a a buffer region, a description of $T$, and three markers. This is necessary because the tape alphabets of $T$ and $U$ need not be identical. The map $\psi$ implements the corresponding decoding.

Note that Eq.~(\ref{eqNotSim}) cannot be used as a definition of the notion of universal TM, unless the map $\psi$ is carefully restricted. In particular, if we only demand that $\psi$ is computable, then the ``identity machine'' that simply outputs the input (i.e.\ that ``does nothing'') would count as universal:
\[
   U(p_T,x)\equiv U(\langle p_T,x\rangle):=\langle p_T,x\rangle.
\]
This is because we can define $\psi$ to be the computable partial function $\psi(\langle p_T,x\rangle):=T(x)$. In this way, we can shift the full computation completely into the process of ``reading the output''. This is certainly not intended.

Thus, we will here use a TM definition which avoids the ``output decoding'' issue:\\
\begin{bozzed}
\begin{definition}[Turing machine (TM)]
\label{Definition:TM}
In the remainder of this paper, unless mentioned otherwise, a TM $T$ is always assumed to conform to the following requirements. The TM has a set of internal states $Q=\{0,1,2,\ldots,k-1\}$ with $k\in\mathbb{N}$, where $q_0:=0$ is the initial state and $q_f:=k-1$ the final state. The TM has two bidirectional tapes (input and work tape), and one unidirectional tape (the output tape). The input tape is read-only, i.e.\ its content cannot be modified during the computation. The finite alphabet for all tapes is $\Sigma=\{0,1,\#\}$. The input is a finite binary string $x\in\{0,1\}^*$ that is initially written on the cells $0,1,\ldots,\ell(x)-1$ of the input tape. All other cells of all tapes are initially blank ($\#$). All tape heads start in position zero. At each step of operation, the input and work tape heads can independently either move to the left or to the right. Furthermore, the machine may write a bit ($0$ or $1$) at the current cell of the output tape and move its output tape head one position to the right (but not to the left), or it leaves the output tape as it is. If the machine halts, i.e.\ enters the distinguished internal state $q_f$, then the TM's output, $y$, is the finite binary string that has so far been written onto the output tape. If this happens, then we write $T(x)=y$. This defines a partial function from the finite binary strings to the finite binary strings.

A \emph{configuration} of a TM is a description of some internal state $q\in Q$, of the number $|Q|$ of internal states, of some positions of the tape heads, and some finite information content of the tapes (all but finitely many cells are blank). The countable set of all possible configurations will be denoted by $\mathcal{C}$.
\end{definition}
\end{bozzed}

By \emph{uni/bidirectional}, we mean that the head can be moved in only one/both of the directions. The notions of input and output are perfectly clear for such a TM. Moreover, a TM of this kind resembles our idea of ``John preparing breakfast'' from Section~\ref{SecIntroduction}: we can observe the machine creating its output bits one after the other. No output bit will ever be erased; the output is complete once the machine has halted. In some sense, the output tape resembles ``John's subsequently observable actions'', and the work tape resembles ``John's brain''.

We can now define a notion of universal TM in the following way.
\begin{bozzed}
\begin{definition}[Universal TM]
\label{Definition:UniversalTM}
A TM $U$ is \emph{universal} if it satisfies the following conditions. There exists a decidable prefix code $\{p_T\}_T$, where $T$ labels the Turing machines, such that $p_T\in\{0,1\}^*$ is a computable description of the TM $T$: that is, there is an algorithm that extracts the set of internal states $Q$ and the transition function $\delta$ from $p_T$. Furthermore,
\begin{equation}
   U(p_Tx)=T(x)\mbox{ for all }x\in\{0,1\}^*;
   \label{eqSim2}
\end{equation}
in particular, $U(p_Tx)$ is undefined if and only if $T(x)$ is undefined (corresponding to the fact the the TM $T$ does not halt on input $x$). Here, $p_T x$ denotes the binary string obtained from concatenating $p_T$ and $x$.
\end{definition}	
\end{bozzed}
That is, a universal TM $U$ takes the description of a TM $T$ as input, and then imitates its output behavior on the rest of the input. Note that this definition is strictly stronger than the usual definition of a universal TM as used in algorithmic information theory~\cite{LiVitanyi,Hutter}: there, it is only demanded that for every TM $T$ there is some $p_T\in\{0,1\}^*$ such that Eq.~(\ref{eqSim2}) holds, but it is not explicitly demanded that a description of $T$ can be reconstructed from $p_T$. Decidability of the prefix code $\{p_T\}$ means that there is a computable function $f:\{0,1\}^*\to\{0,1\}$ with $f(s)=1$ if and only if $s\in\{p_T\}$ --- i.e.\ there exists an algorithm that decides for every given string whether that string is a valid encoding of a TM or not.

Do all universal TMs behave in a way that is similar to Hennie's, i.e.\ as described in Observation~\ref{ObsHennie}? To address this question, we will need a formal definition of a suitable set of ``simple functions'' that extract the configuration of the simulated TM from the configuration of the universal TM. Instead of trying to settle right away how this set $\mathcal{S}$ of simple functions should be defined, let us begin with some basic properties that are immediately clear given our goals. Every $\varphi\in\mathcal{S}$ is a function $\varphi:\mathcal{C}'\to\mathcal{C}'$, where $\mathcal{C}':=\mathcal{C}\cup\{\emptyset\}$ is the set of TM configurations, supplemented by an additional element $\emptyset\not\in\mathcal{C}$ which we can interpret as denoting ``not a valid configuration''. We demand that $\varphi(\emptyset)=\emptyset$ for all $\varphi\in\mathcal{S}$. Furthermore, we assume that $\mathcal{S}$ is closed under composition, i.e.\ if $\varphi,\psi\in\mathcal{S}$ then $\varphi\circ\psi\in\mathcal{S}$, and that the identity map is in $\mathcal{S}$. We also assume that the functions in $\mathcal{S}$ are computable and total.

Now consider the sequence of configurations $\mathcal{C}_U(p_T x,t)$ for $t=0,1,2,\ldots$. The idea is that for some $t'$, the universal TM $U$ has just completed another simulated step of operation of $T$. If we denote the number of simulated time steps by $t$ (such that $t\leq t'$), then we would like our simple function $\varphi$ to yield
\[
   \mathcal{C}_T(x,t)=\varphi\left(\mathcal{C}_U(p_T x,t')\right).
\]
Otherwise, the right-hand side will simply yield $\emptyset$. On the one hand, this means that we obtain
\[
   \mathcal{C}_T(x,t)=\varphi\left(\mathcal{C}_U(p_T x,\tau_T(t))\right)
\]
exactly as in Observation~\ref{ObsHennie}, for some increasing set of integers $t\mapsto \tau_T(t)$. On the other hand, if $T$ halts on input $x$ at time $t_H$, this allows us to obtain the sequence
\[
   \mathcal{C}_T(x,0),\mathcal{C}_T(x,1),\ldots,\mathcal{C}_T(x,t_H)
\]
by parsing the sequence
\[
   \mathcal{C}_U(p_T x,0),\mathcal{C}_U(p_T x,1),\ldots,\mathcal{C}_U(p_T x,t_H')
\]
via application of $\varphi$ to each entry, discarding those where $\varphi$ takes the value $\emptyset$. This motivates the following definition.
\begin{bozzed}
\begin{definition}[Simulation preorder]
\label{DefSimulationPreorder}
Let $T$ and $T'$ be TMs. Suppose that, relative to our choice of simple functions $\mathcal{S}$, there is some $\varphi\in\mathcal{S}$ such that for every input $x\in\{0,1\}^*$, the sequence of configurations
\[
   \mathcal{C}_T(x,0),\mathcal{C}_T(x,1),\ldots,\mathcal{C}_T(x,t_H),
\]
where $t_H$ is the halting time if $T$ halts on input $x$, and $\infty$ otherwise, can be obtained by processing the sequence of configurations
\[
   \mathcal{C}_{T'}(x,0),   \mathcal{C}_{T'}(x,1),    \mathcal{C}_{T'}(x,2),\ldots,   \mathcal{C}_{T'}(x,t_H'),
\]
where $t'_H$ is the halting time of $T'$ on input $x$, and $\infty$ otherwise, in the following way. One after the other, apply $\varphi$ to the $\mathcal{C}_{T'}(x,\bullet)$, and if the result is not $\emptyset$, then append the result to the list. In this case, we will say that \emph{$T'$ simulates $T$} and write
\[
   T\preceq_{\mathcal{S}} T'.
\]
Note that this implies that $t_H\leq t'_H$.
\end{definition}
\end{bozzed}
Since the identity function is in $\mathcal{S}$, we have $T\preceq_{\mathcal{S}} T$ for all TMs $T$. Furthermore, $T\preceq_{\mathcal{S}} T'$ and $T'\preceq_{\mathcal{S}} T''$ implies $T\preceq_{\mathcal{S}} T''$, since simple functions can be composed. This implies that $\preceq_{\mathcal{S}}$ is a \emph{preorder} on the TMs.

The notion of a simulation preorder is well-known in the literature. However, all definitions that we are aware of, including Milner's seminal work~\cite{Milner}, define a notion of simulation that is too strict for our purpose. To the best of our knowledge, these definitions postulate that some machine $S$ simulates another machine $T$ if there is an injection of $T$'s states into those of $S$ such that each single state transition of $T$ corresponds to a single state transition of $S$. These definitions are very natural for finite automata~\cite{Park}. However, for TMs, we need a looser definition of simulation that allows $S$ to simulate a single step of $T$ within more than one time step.

When we compare the computation of $T$ with its simulation by $U$, we have an additional prefix $p_T$ on the input, as explained above. The behavior sketched above will then be abbreviated by writing $T\preceq_{\mathcal{S}} U(p_T\bullet)$. We use this notation for our first attempt to formalize our conjecture:
\begin{bozzed}
\textbf{Conjecture (1st attempt).} For every universal TM $U$, we have
\begin{equation}
   T\preceq_{\mathcal{S}} U(p_T\bullet)\mbox{ for every TM }T.
   \label{eqSim3}
\end{equation}
\end{bozzed}
At first sight this seems plausible: universal TMs operate by taking the description $p_T$ of any other TM $T$, and by then simulating $T$ step by step.

However, it is easy to see that this conjecture cannot literally hold true. Consider a Turing machine $T_0$ that operates as follows. On input $x$, it begins by computing the $2^{\ell(x)}$th prime number in binary on its work tape. After that, it halts unconditionally. This Turing machine $T$ will output the empty string on every input, i.e.\ $T_0(x)=\varepsilon$ for all $x\in\{0,1\}^*$, and it will do so extremely inefficiently. In fact, let us consider an infinite sequence of modifications of this inefficient machine, labelled by $T_i$. All the $T_i$ are identical to $T_0$, but they have $i$ additional internal states that are all irrelevant for all their computations. That is, the transition functions of all $T_i$ are that of $T_0$, and if $Q_0=\{0,1,2,\ldots,k-1\}$ is the set of internal states of $T_0$, then the set of internal states of $T_i$ is $Q_i:=\{0,1,2,\ldots,k-1,k,\ldots,k-1+i\}$.

Consider now some standard textbook universal TM $U$, but modify it such that it does the following. In the very beginning, $U$ will examine the $p_T$-part of the input, and check whether $p_T$ is a description of any of the $T_i$, i.e.\ whether $p_T=p_{T_i}$ for some $i$. If not, then $U$ will proceed like the textbook machine; otherwise, it will refrain from simulating $T_i$, and instead halt (and output $\varepsilon$) immediately and unconditionally.

In this case, there will be infinitely many counterexamples to Eq.~(\ref{eqSim3}), namely all the $T=T_i$ will violate it.

Indeed, we have already formulated our Informal Conjecture in a more careful way: not all TMs $T$, but only those that lie in a large set $\mathbf{T}$ are conjectured to represent instances of computational sourcehood. Intuitively, $\mathbf{T}$ contains all TMs that operate neither in a trivial nor in an extremely inefficient way on all inputs. This leads us to the second attempt at formalization of our Information Conjecture:
\begin{bozzed}
\textbf{Conjecture (2nd attempt).} For every universal TM $U$, we have
\[
   T\preceq_{\mathcal{S}} U(p_T\bullet)\mbox{ for every TM }T\in\mathbf{T},
\]
where $\mathbf{T}$ is a large set of TMs yet to be formalized but intuitively described above. Moreover, if we denote the simple function that implements the simulation decoding of $T$ by $\varphi_T$, then the map $p_T\mapsto\varphi_T$ is computable.
\end{bozzed}

We have added another desideratum to the conjecture: that given $T$'s description $p_T$, a finite algorithm can actually determine an effective description of the simple function $\varphi_T$ which reads out $T$'s simulated configuration from $U$'s. We assume that this algorithm yields some valid $\varphi_T$ for all $p_T$, but the result must only be correct for $T\in\mathbf{T}$. This demand is very natural: not only would we like $T$'s computation to be \emph{in principle} determinable from $U$'s, but it \emph{should generally be known how to actually do so}.

If the set of simple functions $\mathcal{S}$ is closed under certain operations, we can equivalently demand that the decoding function is independent of $T$, i.e.\ ``one simple function reads them all'':
\begin{bozzed}
\begin{lemma}
Suppose that the set of simple functions $\mathcal{S}$ has the following \emph{prefix closure property}: for every decidable prefix code $\{p\}$ labelling a subset of the simple functions $\{\varphi_p\}\subseteq\mathcal{S}$, the function $\varphi:\mathcal{C}'\to\mathcal{C}'$,
\[
   \varphi(c):=\left\{
      \begin{array}{cl}
      	\varphi_p(c) & \mbox{ if input tape of }c\mbox{ starts with }p\\
      	\emptyset & \mbox{ otherwise},
      \end{array}
   \right.
\]
is also contained in $\mathcal{S}$. Then we can without loss of generality assume that for every universal TM $U$, there is a \emph{unique} simple function $\varphi\in\mathcal{S}$ that implements the simulation decoding for \emph{every} TM $T$.
\end{lemma}	
\end{bozzed}
Next we will discuss how to concretely choose a suitable set of simple functions $\mathcal{S}$.

\subsection{How not to choose the set of simple functions $\mathcal{S}$}
It is clear that \emph{every} rigorous formulation of our conjecture must be false if the set of simple functions $\mathcal{S}$ is ``too small''. For example, suppose that we choose $\mathcal{S}=\{{\rm id}\}$, i.e.\ define only the identity function to be simple. In this case, $T\preceq_{\mathcal{S}} U(p_T\bullet)$ implies
\[
   \mathcal{C}_T(x,t)=\mathcal{C}_U(p_Tx,t)\quad\mbox{for all } x\in\{0,1\}^*\mbox{ and all }t\in\mathbb{N}.
\]
But this is certainly impossible (unless $T=U$ and $p_T=\varepsilon$): already at $t=0$, the input tape contents of $T$ and $U$ differ (they hold the strings $x$ and $p_T x$ respectively), hence $T$ and $T'$ have different configurations, and $\mathcal{C}_T(x,0)\neq\mathcal{C}_U(p_T x,0)$.

More generally, the following holds:
\begin{bozzed}
\begin{lemma}
If we define $\mathcal{S}$ to be minimal, i.e.\ to only contain the identity function ($\mathcal{S}=\{\mathrm{id}\}$), then
\[
   T\preceq_{\mathcal{S}}T'\Leftrightarrow \mathcal{C}_T(x,t)=\mathcal{C}_{T'}(x,t)\mbox{ for all }x,t.
\]
That is, TMs formally only ever simulate machines that are exactly identical to themselves in all their state transitions. In particular, $T\preceq_{\mathcal{S}}T'$ becomes equivalent to $T'\preceq_{\mathcal{S}}T$.
\end{lemma}	
\end{bozzed}
Note that the right-hand side is not the same as $T=T'$: for example, $T=T_i$ and $T'=T_j$ for $i\neq j$ from the family of inefficient TMs under Eq.~(\ref{eqSim3}) will also satisfy it. However, TMs $T$ and $T'$ that satisfy the right-hand side above are ``identical for all practical purposes''.

A similar conclusion will follow if we choose the set $\mathcal{S}$ non-trivial but still too small: for our conjecture to be true, $\mathcal{S}$ must contain all possible ways in which universal computers can choose to encode the simulation in their own configuration. In particular, $\mathcal{S}$ must at least contain all ``textbook simulation encodings'', like the one used by Hennie as described in Subsection~\ref{SubsecTextbook}.

On the other hand, suppose we define $\mathcal{S}$ to be the set of \emph{all} total computable functions $\varphi$ with $\varphi(\emptyset)=\emptyset$. Then this will make our conjecture trivially true for \emph{many} universal computers $U$, but it will in general fail to formalize a sensible notion of simulation, as we will now demonstrate.

For the sake of the argument, let us assume that there exists some universal TM $V$ that intuitively violates our Informal Conjecture: it reproduces the outputs of all (or most) other TMs $T$ in a counterintuitive way that is very different from step-by-step simulation. Let us construct another universal TM $U$ with $U(x)=V(x)$ for all $x\in\{0,1\}^*$ --- it will be a machine that we obtain by modifying $V$, and that \emph{also} violates our Informal Conjecture. It is constructed in the following way. To obtain the required output functionality, $U$ simulates $V$ exactly step by step. In addition, $U$ contains a \emph{counter} on some unused portion of its work tape, i.e.\ a representation of a natural number $\tau$ that starts in zero and increases by one after every step of simulation of the TM $V$. Now we define a total computable function $\varphi$ via the following algorithm:
\begin{itemize}
	\item Extract $p_T$ and $x$ from the input tape and $t$ from the work tape.
	\item Simulate $T$ on input $x$ for $t$ steps and return the configuration $\mathcal{C}_T(x,t)$.
\end{itemize}
Consequently, we obtain $T\preceq_{\mathcal{S}} U(p_T\bullet)$ for \emph{all} TMs $T$. However, by construction, $U$ never \emph{actually} performs any step-by-step simulation of any TM $T$ (since $V$ doesn't). We have thus shown the following undesirable feature of the maximal choice of $\mathcal{S}$ as the set of all total computable functions: \emph{if} there exist universal TMs that violate our Informal Conjecture, \emph{then} some of them will still satisfy $T\preceq_{\mathcal{S}} U(p_T\bullet)$. Thus, $\prec_{\mathcal{S}}$ is not a realiable formalization of the notion of step-by-step simulation that our Informal Observation refers to.

To expose the problem further, consider the following special TM:
\begin{bozzed}
\begin{definition}[Clock Turing Machine]
\label{DefClockTM}
A \emph{clock Turing machine} $C$ is a TM that ignores its input and counts integer time steps $t\in\N$ on its work tape indefinitely.
\end{definition}
\end{bozzed}
We are not giving a formal construction of a clock TM, but it is not difficult to think of a concrete set of internal states and a transition function that implements the clock. For example, at each time step, the TM may simply write a fixed symbol (say, $1$) in the currently active cell of the work tape and move the work tape head to the right. Since this can be done in different ways (e.g.\ writing only zeros or ones, or alternating in ways that are determined by changes of the internal state), there are infinitely many clock TMs. Clock TMs $C$ never halt, i.e.\ $C(x)$ is undefined for every $x\in\{0,1\}^*$.

The construction above shows the following:
\begin{bozzed}
\begin{lemma}
\label{LemClock}
Let $C$ be a clock TM. If we define $\mathcal{S}$ to be maximal, i.e.\ equal to the set of all total computable functions $\varphi$ with $\varphi(\emptyset)=\emptyset$, then
\[
   T\preceq_{\mathcal{S}}C\quad\mbox{for every TM }T.
\]
That is, the clock TM $C$ will formally be considered to simulate all other TMs step by step.
\end{lemma}	
\end{bozzed}
Most total computable functions are intuitively \emph{extremely complex}, so the maximal choice of $\mathcal{S}$ is obviously a very bad formalization of a ``set of simple functions''. However, the argument above rules out other, more intuitively sensible choices of $\mathcal{S}$. For example, we may consider the set $\mathcal{S}$ of functions that have at most linear time complexity. A running time at least linear in the input length is required to read the input, and as we would like the functions in $\mathcal{S}$ to be simple, it is natural to demand that they shall not take significantly more time than this minimum.

However, we will now argue that even such functions with linear running time can still be too powerful, as they also allow to apply the clock TM trick. Since we are considering functions on configurations, we have to be more specific about what we mean by linear time complexity.  
	For our purpose, we will restrict our attention to the following specific choice of encoding of configurations and TMs $M_\varphi$ that compute $\varphi \in \mathcal{S}$. The TM $M_{\varphi}$ is supposed to satisfy Definition~\ref{Definition:TM}, up to a convenient modification: instead of \emph{single} input, work, and output tapes, $M_{\varphi}$ has two input tapes ($I_{\rm in}$ and $I_{\rm out}$), two output tapes ($O_{\rm in}$ and $O_{\rm out}$), and four work tapes ($W_{\rm in}$, $W_{\rm out}$, $W$, and $W_{\rm state}$).

We use the convention that $M_{\varphi}$ receives the input configuration $c\in\mathcal{C}$ in the following way. The tapes and the corresponding head positions will exactly be copied onto the in-tapes. That is, $I_{\rm in}$ contains the exact input tape content as described by $c$, and also its tape head will be placed at the current position specified by $c$; similarly for $W_{\rm in}$ and $O_{\rm in}$. The internal state $q\in Q$ described by $c$ will be written via $\lceil \log_2 |Q|\rceil$ bits onto the tape $W_{\rm state}$.

When $M_{\varphi}$ has halted, the tapes $I_{\rm out}$, $W_{\rm out}$ and $O_{\rm out}$ contain the respective tape contents and tape head positions of the target configuration $c'=\varphi(c)$, and the tape $W_{\rm state}$ contains the description of the internal state of $c'$. If the content of $W_{\rm state}$ is not a syntactically correct description of any $q\in Q$, then the output configuration is taken to be $\emptyset$.

Given such $\varphi$, let us say it has \emph{linear time complexity} or \emph{linear running time} if the number of computation steps of $M_\varphi$ is in $\mathcal{O}(n)$, where $n$ is the total size of the non-blank blocks on the initial configuration's input, work, and output tapes.

We will now show that even those functions are too powerful to represent a meaningful notion of simple functions. Consider again a clock TM $C$ of Lemma~\ref{LemClock}. Given a TM $T$ that is supposed to be simulated, we can construct a ``simple'' function $\varphi_T$ via some TM $M_{\varphi_T}$ as follows. The TM $M_{\varphi_T}$ has a set of internal states $Q=Q_T\times Q'$, where the $Q'$ part is used to carry its functionally relevant internal state, while $Q_T$ carries the representation of an internal state of $T$. First, $M_{\varphi_T}$ copies the input of $T$ from $I_{\rm in}$ to $I_{\rm out}$. Then $M_{\varphi_T}$ sets the first component of its internal state equal to the starting state $q_0\in Q_T$ of $T$. From that moment on, $M_{\varphi_T}$ behaves exactly like $T$, with $I_{\rm out}$ as its input tape, $W_{\rm out}$ as its (initially blank) work tape, $O_{\rm out}$ as its (initially blank) output tape, and $Q_T$ as state register: we assume that the transition function of $M_{\varphi_T}$ contains the transition function of $T$ and uses it for an exact step-by-step simulation of $T$. After every simulated time step, $M_{\varphi_T}$ moves the head on $W_{\rm in}$ to the left. When that head reaches a blank symbol, $M_{\varphi_T}$ represents the first part of its internal state via $\lceil\log_2|Q_T|\rceil$ bits on the tape $W_{\rm state}$ and halts. 

The TM $M_{\varphi_T}$ operates in linear time and maps configurations to configurations. If we apply $M_{\varphi_T}$ to the clock TM $C$, then $M_{\varphi_T}$ will simulate $T$ for as many steps $t$ as there are ones on $C$'s work tape. This clearly produces the configuration $\mathcal{C}_T(x,t)$ in linear time. Thus, $\varphi_T\in\mathcal{S}$, and we have shown the following:
\begin{bozzed}
\begin{lemma}
\label{LemClockLinear}
Let $\mathcal{S}$ be the set of total computable functions (on the configurations) that run in \emph{linear time}, in the sense explained above. Then we still have
\[
   T\preceq_{\mathcal{S}}C\quad\mbox{for every TM }T
\]
if $C$ is any clock TM.
\end{lemma}	
\end{bozzed}
Can stricter time bounds give us a better choice of $\mathcal{S}$? This seems unlikely, given that at least linear time is needed to even \emph{read} the input configuration. So is there another way to define a set of simple functions $\mathcal{S}$ that gives us a nontrivial simulation preorder, but that leaves some chance for our conjecture (say, in its 3rd formalization) to be true? Unfortunately, there is a strong counterargument to this hope, as we will now demonstrate.

\subsection{An encryption counterexample to ``simplicity''}
Let us now show that our conjecture cannot hold without substantial modification. Since a detailed formal proof of what follows seems very cumbersome, we will instead give an informal argument which we believe is sufficiently detailed to support our conclusion.

Consider some universal TM $U$ of the textbook kind --- similarly as Hennie's universal TM, $U$ simulates every other TM $T$ step by step. Let us construct a modification of $U$, denoted $U'$, that has equivalent input-output behavior as $U$, i.e.\ $U(x)=U'(x)$ for all binary strings $x$. However, $U'$ will encrypt all the parts of $T$ that are not relevant for the current simulation step. Then functions $\varphi \in \mathcal{S}$ that deserve the name ``simple'' will not be able to crack the encryption of $U'$s representation of $T$.

To this end, consider some computable function that maps integers $n\in\N$ to bits $a_n\in\{0,1\}$. Let us choose a function that is very difficult to evaluate: For example, we may assume that it takes exponentially many time steps to compute $a_n$. More specifically, the time hierarchy theorems~\cite{AroraBarak} guarantee the existence of a decision problem that cannot be solved within time $\mathcal{O}(2^n)$, but within time $\mathcal{O}(2^{2n})$, and we choose $a_n$ to be the answer to such a problem. For negative $i$, we will use the convention $a_i:=a_{|i|}$. We will use this string of bits for encryption.

Like $U$, the TM $U'$ simulates all elements of $T$ step by step. In particular, it contains the contents $\{w_i\}_{i\in\mathbb{Z}}$ on $T$'s work tape cells somewhere in its memory. While simulating a single time step of $T$, the TM $U$ will read the current tape cells, determine and write their new contents, and then move the simulated tape heads left or right. We construct $U'$ such that it replaces this one simulated step of operation of $T$ by the following simulation steps (all tape cells are \emph{simulated} tape cells):
\begin{itemize}
	\item It reads the symbol in the currently active work tape cell $i$ (assumed unencrypted) and the other currently active tape cells and applies $T$'s tabulated transition function $\delta$ to determine whether it has to move left or right on the tapes, and which symbol $w_i$ it has to write into the work tape cell (and similarly for the other tapes).
	\item Then it computes, with a fixed program independent of any other tape content, the bits $a_i$ and $a_{i+\sigma}$, where $\sigma=-1$ if it has to turn left on the work tape or $\sigma=+1$ if it has to turn right.
	\item It determines $w_i'$, which is the blank symbol $\#$ if $w_i$ is blank, and which is $w_i\oplus a_i$ if $w_i$ is a bit (encryption). It writes $w'_i$ into work tape cell $i$ and reads $w'_{i+\sigma}$ from work tape cell $i+\sigma$. It then determines $w_{i+\sigma}$, which is the blank symbol $\#$ if $w'_{i+\sigma}$ is blank, and  $w'_{i+\sigma}\oplus a_{i+\sigma}$ otherwise (decryption). The writing onto the input and output tape is performed without encryption, as determined by $T$'s transition function $\delta$.
	\item $U'$ erases all data that results from the computation of $a_i$ and $a_{i+\sigma}$ and of the sums of those with the work tape bits from other parts of its memory.
\end{itemize}
We assume that the TM $U'$ is constructed such that the only relevant difference after the encryption resp.\ decryption step to before is the value of the simulated work cell bit. In other words, $U'$ is supposed to ``erase all the garbage'' that it produced while computing $a_i$, leaving only a simple encoding of the encrypted configuration of $T$. We also assume that $U'$ does \emph{not} contain an explicit counter of the number of time steps that have passed since the start of the computation.

Let $s_1,s_2,s_3,\ldots$ label the times at which the TM $U'$ has completed simulating (as described above) the first, second, third$\ldots$ step of computation of $T$. Now suppose that the current formalization of our conjecture (2nd attempt) is true, for some intuitively reasonable set of simple functions $\mathcal{S}$. Then there is some $\varphi'\in\mathcal{S}$ and an increasing sequence of times $\{\tau'_t\}_t$ such that
\begin{equation}
   \varphi'(\mathcal{C}_{U'}(p_T x,\tau'_t))=\mathcal{C}_T(x,t)
   \label{eqEncrRead}
\end{equation}
for all $t$, and such that $\varphi'(\mathcal{C}_{U'}(p_T x,s))=\emptyset$ for all $s\not\in\{\tau'_t\}_t$. A priori, the times $\tau'_t$ need not at all be related to the simulation times $s_t$. For example, $\varphi'$ could act similarly as the function that we have used to prove Lemma~\ref{LemClock}: it could simply perform the simulation of $T$ itself, without having to wait for $U'$ to have finished its simulation steps. Then we could have, for example, that $\tau'_t=t$, but $s_t$ will grow exponentially with $t$.

However, such functions $\varphi$ are not of the form that we have in mind in any intuitive formulation of our conjecture. Let us therefore assume that the set of simple functions $\mathcal{S}$ \emph{does} comply with out intuition to some extent: \emph{the simple functions shall not simulate $T$ by themselves, but shall only read was has already been simulated by the universal TM $U'$}. Intuitively, the simulated configuration $\mathcal{C}_T(x,t)$ is generated (in encrypted form) at time $s_t$ on $U'$ and not before. Thus, our assumption amounts to postulating that the times $\tau'_t$ must essentially be on or after the times $s_t$ when $U'$ has performed the $t$th step of simulation of $T$:

\textbf{Assumption 1.} There is a possible choice of $\varphi'$ such that there exists at least one $x$ such that $s_t\leq\tau'_t<s_{t+1}$ for many different $t$.

Note that we are not assuming that this has to hold for \emph{all} TMs $T$. We only need to assume that there exists at least one TM $T$ within the ``sufficiently diverse'' set of TMs (as mentioned in the 2nd attempt of formalization of our conjecture) that satisfies this assumption and the other two below.

In other words, for many $t$, the simple function $\varphi'$ has to read out $\mathcal{C}_T(x,t)$ within the time interval $[s_t,s_{t+1}-1]$ --- this may not be true for \emph{all} simple functions that read out $T$'s configuration, but we assume that it is true for \emph{some} $\varphi'$.

Now, via Eq.~(\ref{eqEncrRead}), let us analyze what this means for such $\varphi'$. Consider one of the ``many different $t$'' from Assumption 1. Then the simple function $\varphi'$ has to determine $\mathcal{C}_T(x,t)$ from some $\mathcal{C}_{U'}(p_T x,s)$, where $s_t\leq s < s_{t+1}$. What do these configurations of $U'$ look like? For $s=s_t$, this configuration consists of a simple encoding of $\mathcal{C}_T(x,t)^*$, by which we denote $\mathcal{C}_T(x,t)$ with \emph{all work tape bits but one} encrypted. This is basically it --- by construction, $U'$ does not contain more information than that (except for, say, a constant set of instructions that allows $U'$ to compute any $a_i$ or to decode the transition function $\delta_T$ from $p_T$ etc.).

Over the next time steps, for $s_t<s<s_{t+1}$, the TM $U'$ computes $a_i$ and $a_{i+\sigma}$ and uses this to encode and to decode the currently and next active cell on $T$'s simulated work tape. Over all these time steps, $U'$ will thus contain a simple encoding of $T$'s configuration with \emph{all work tape bits but at most two} encrypted --- let us denote this by $\mathcal{C}_T(x,t)^ {**}$ --- and some ``rest''. This ``rest'' contains, for example, intermediate results of the computation of $a_i$ and $a_{i+\sigma}$, but it does \emph{not} contain any information of substantial help in decrypting any of the other work tape bits. Schematically, we may thus write
\[
   \mathcal{C}_{U'}(p_T x,s)=\left(\strut \mathcal{C}_T(x,t)^{**},{\rm rest}\right),
\]
and the above remarks motivate us to formulate the following

\textbf{Assumption 2.} Computing $\mathcal{C}_T(x,t)$ from $\mathcal{C}_{U'}(p_T x,s)$ is essentially equivalent to computing it from $\mathcal{C}_T(x,t)^{**}$ --- in this sense, the ``rest'' can be ignored.

But now, recalling Eq.~(\ref{eqEncrRead}), we find that $\varphi'$ essentially has to \emph{map $\mathcal{C}_T(x,t)^{**}$ onto $\mathcal{C}_T(x,t)$} --- it seems like this means that $\varphi'$ has to decrypt all but two of $T$'s work tape bits, which is a Herculean task if the work tape contains many bits.

Or does it really have to? Note that there may be \emph{indirect} ways of determining $\mathcal{C}_T(x,t)$ from $\mathcal{C}_T(x,t)^{**}$ without actually doing any decryption. For example, suppose that $T=C$ is the clock TM from Definition~\ref{DefClockTM}. Then $\mathcal{C}_T(x,t)$ has a sequence of $t$ identical ones on its work tape (followed by blank tapes). This configuration can be determined by simply counting how many non-blank symbols are on the work tape of $\mathcal{C}_T(x,t)^{**}$, and then by replacing each bit by a one.

However, the clock TM is very special. In contrast, think of a TM $T$ with the property that the configuration $\mathcal{C}_T(x,t)$ differs, for a large set of values of $t$ (and identical $x$), only in the content of the work tape bits. (For example, $T$ could simply count integers in binary on the work tape for a very large number of time steps, without modifying other parts of its configuration.) Then, the only way to determine $\mathcal{C}_T(x,t)$ from $\mathcal{C}_T(x,t)^{**}$ would be by brute-force decryption.

\textbf{Assumption 3.} $T$ writes many bits onto its work tape \emph{and} has the property that $\mathcal{C}_T(x,t)$ can, for many $x$ and $t$, essentially only be determined by brute force decryption of $\mathcal{C}_T(x,t)^{**}$.

Then we obtain the following consequence.\\ \\
\begin{bozzed}
\begin{observation}
\label{Obs10}
If the above informal argumentation can be made rigorous (including in particular the three assumptions), then our conjecture (in its 2nd attempt of formalization) can only be correct if we allow the set of simple functions $\mathcal{S}$ to contain maps that are intuitively extremely complex.
\end{observation}	
\end{bozzed}

\subsection{From simplicity to preservation of structure}
Our goal was to find a formalization of the idea that \emph{all functionally relevant elements of $T$ must typically be fully represented within $U$} --- a formulation that is non-trivial and has a chance to be true. But it seems that we have tried to do so in the wrong way: demanding that $T$'s configuration can be read from $U$'s with some function that is \emph{simple to implement} is doomed to fail --- at least if we understand simplicity as small time complexity. On the one hand, the Observation~\ref{Obs10} suggests that $\mathcal{S}$ must necessarily contain very complex functions to break the encryption of some universal TMs. On the other hand, Lemma~\ref{LemClockLinear} shows that even very small time complexity already admits functions that ``cheat'' by performing the simulation themselves.

Note that putting a time bound on admissible read-out functions is similar to the strategy by Zwirn and Delahaye~\cite{Zwirn1,Zwirn2} to define ``approximations'' of Turing machines in their approach to construct a formal definition of computational irreducibility. The above shows that our approach to define computational sourcehood (and its typicality) needs a different strategy.

For a different perspective, consider again the example of the clock TM of Lemma~\ref{LemClockLinear}. Why should we regard functions $\varphi\in\mathcal{S}$ that implement the simulation by themselves as ``undesired'' or ``cheating''? Our previous attempt was to say that such $\varphi$ are not \emph{simple} in any meaningful sense of the word. But there is an alternative view: we can also say that such $\varphi$ are \emph{not sufficiently structure-preserving}.

To see this, let us contrast such ``cheating'' $\varphi$ with typical read-out functions $\psi$ for textbook universal TMs like Hennie's. Consider two configurations $c,c'\in\mathcal{C}$ of a simulated TM $T$ that are in some sense ``pretty close to each other'' --- perhaps they differ only in a small number of bits on the work tape, but are otherwise identical. Consider some configuration $c_U\in\mathcal{C}$ which describes the universal TM $U$ simulating $T$ in configuration $c$, i.e.\ $\psi(c_U)=c$. Then we will find another configuration $c'_U$ \emph{close to $c_U$} which describes $U$ simulating $T$ in configuration $c'$, i.e.\ $\psi(c'_U)=c'$: intuitively, we only have modify a few bits of $c_U$ (those that represent the simulated bits differing between $c$ and $c'$) to obtain $c'_U$ from $c_U$.

In contrast, the ``cheating'' $\varphi$ for the clock TM $C$ will not in general satisfy this property: there will be close-by configurations $c=\mathcal{C}_T(x,t)$ and $c'=\mathcal{C}_T(x',t')$ with $c=\varphi(c_C)$ such that \emph{all} configurations $c'_C$ with $c'=\varphi(c'_C)$ are very far away from $c_C$. To see this, let $T$ be a universal TM. For every input $x\in\{0,1\}^*$, denote by $t_H(x)$ $T$'s halting time on input $x$ (which is $\infty$ if it does not halt on that input). Furthermore, for those $x$ with $t_H(x)<\infty$, denote by $N(x)$ the smallest $i\in\N$ such that the $i$th output tape cell is blank at halting; in other words, this means that the TM halts with a block of $N(x)$ non-blank bits on its output tape; for non-halting $x$, set $N(x):=-1$. Then $N_n:=\max_{x\in\{0,1\}^n} N(x)$ grows extremely fast --- similarly as the busy beaver function~\cite{Rado}, it must grow faster than every computable function, due to the undecidability of the halting problem. For every $n$, pick an arbitrary maximizing input $x_n$, i.e.\ $N(x_n)=N_n$ and $\ell(x_n)=n$. Then the configuration $c_n:=\mathcal{C}_T(x_n,t_H(x_n))$ contains an extremely large (of size $N_n$) block of bits on its output tape. For every $i\in [0,N_n]$, denote by $c_n^{(i)}$ the configuration $c_n$ with the $i$th bit on the output tape inverted. In other words, $c_n$ differs from every $c_n^{(i)}$ in only a single bit. Now, by definition of $\varphi$ and of the clock TM,
\[
   c_n=\varphi(\mathcal{C}_C(x_n,t_H(x_n)))=\varphi(c_{n,C}),
\]
where $c_{n,C}:=\mathcal{C}_C(x_n,t_H(x_n))$. Let us estimate the number of configurations $c'_{n,C}$ that have distance at most $k$ from $c_{n,C}$, where $k\in\N\setminus\{0\}$, i.e.\ the number of elements in the $k$-ball $B_k(c_{n,C})$. We have not formally defined a distance measure on the configurations yet, and our argumentation will not be particularly sensitive to the choice of measure. Nonetheless, for concreteness, let us define
\[
   D(\mathcal{C}_C(x,t),\mathcal{C}_C(x',t')):=D_H(\bar x,\bar x')+|\ell(x)-\ell(x')|+|t-t'|,
\]
where $\bar x$ and $\bar x'$ denote the first $m$ bits of $x$ resp.\ $x'$, where $m:=\min\{\ell(x),\ell(x')\}$, and $D_H$ is the Hamming distance. (All other configurations will be mapped to $\emptyset$ by $\varphi$, hence we are not interested in them). Clearly, $D(\mathcal{C}_C(x',t'),c_{n,C})\leq k$ implies $|t'-t_H(x_n)|\leq k$ and $\ell(x')\leq \ell(x)+k$. There are $(2k+1)$ many choices of such $t'$ and $2^{n+k}$ many choices of such $x'$. Since $2k+1< 2^{2k}$,
\[
   |B_k(c_{n,C})|< 2^{n+3k}.
\]
Now let $k_n:=\lfloor \frac 1 3 (\log_2 N_n-n)\rfloor$, which still grows faster in $n$ then every computable function. By simple counting, there must be at least one $i\in[0,N_n]$ such that $\varphi(\tilde c)\neq c_n^{(i)}$ for every $\tilde c \in B_k(c_{n,C})$: there are simply not enough configurations in the $k$-ball to cover all $c_n^{(i)}$. Hence for every $n$ there exists some $i$ such that $D(c_n,c_n^{(i)})=1$ and $c_n=\varphi(c_{n,C})$, but
\[
   c_n^{(i)}=\varphi(c_{n,C}^{(i)})\Rightarrow D(c_{n,C}^{(i)},c_{n,C})> k_n,
\]
where $k_n$ grows extremely quickly in $n$. Hence $\varphi$ is not structure-preserving in the way explained above.

This motivates our final attempt of formalizing our conjecture:
\begin{bozzed}
\textbf{Conjecture (3rd and final attempt).} For every universal TM $U$, we have
\[
   T\preceq_{\mathcal{S}} U(p_T\bullet)\mbox{ for every TM }T\in\mathbf{T},
\]
where $\mathcal{S}$ is a natural set of structure-preserving maps on TM configurations.
\end{bozzed}
By structure-preserving maps, as sketched above, we mean functions $\varphi$ with the following property. If we have close-by configurations $c$ and $c'$, and another configuration $c_U$ with $\varphi(c_U)=c$, then there is another configuration $c'_U$ \emph{close to $c_U$} with $\varphi(c'_U)=c'$. These functions are not necessarily assumed to be ``easy to implement''.

It is not clear whether (and perhaps even unlikely that) the best definition of ``close-by'' is similar to the one used above, i.e.\ based on a distance measure between configurations that is essentially some modification of the Hamming distance. It may well be that it is more suitable to introduce a form of ``functional similarity'', perhaps a notion that is allowed to depend on the Turing machine under consideration. It is hence likely that higher-level mathematical tools, perhaps from category theory, are needed to substantiate this attempt. Whether the above attempt can be made rigorous in some such way will have to be seen in future work.

\section{Conclusions}
\label{SecConclusions}
In this article, we have revisited the idea that computer science notions like computational irreducibility~\cite{Wolfram} can shed light on the relation between determinism and free agency~\cite{SEPFreeWill}. We have addressed two issues with previous proposals of this kind: first, Wolfram's original proposal did not include a rigorous mathematical definition of computational irreducibility, and it is unclear whether later definitions, such as those by Zwirn and Delahaye~\cite{Zwirn1,Zwirn2}, are well suited to reason specifically about free agency. Second, as argued also by Bringsjord~\cite{Bringsjord}, the focus of Wolfram's (and Lloyd's~\cite{Lloyd}) approach on the question of \emph{temporal shortcuts} allows us to reason about \emph{unpredictability as a phenomenon} of free agency, but not directly about the question of whether agents' decisions are \emph{actually} free.

Motivated by a simple thought experiment (``John the cook'', cf.\ Figure~\ref{fig_john}), we have proposed a variant of computational irreducibility, termed \emph{computational sourcehood}, that is intended to formalize an aspect of actual free agency more directly. We suggest that a process $P$ can be regarded as the \emph{source} of its outputs if attempts to reproduce them must typically involve a step-by-step simulation that contains replicas of the history of configurations of $P$. While this notion is closely related to computational irreducibility (in particular, it also implies that there are typically no shortcuts to simulating $P$), it makes a more general claim, by stipulating that even slow and inefficient simulations must typically contain ``clones'' of the process.

We have then taken up the challenge to give a rigorous mathematical formulation of this phenomenon and its conjectured typicality. This has led us to a question about universal Turing machines (TMs), defined by their ability to reproduce the input-output behavior of all other TMs $T$: \emph{is it true that all universal TMs work by essentially simulating $T$ step by step, except for a small subset of TMs $T$ for which they know shortcuts?} While this question -- and the conjecture of a positive answer to it -- can easily be described in words, it turns out to be quite difficult to find a rigorous formulation that is non-trivial and has a chance to be true.

Our first idea of formalization was to say that $U$ simulates $T$ step by step if the temporal sequence of configurations of $T$ can be read out from the sequence of configurations of $U$ via some ``simple function''. We have shown that this is true for textbook constructions of universal TMs like Hennie's~\cite{Hennie}, but we have identified obstructions to proving that this must be true in general. On the one hand, we have sketched a universal TM that encrypts most of the simulated data, which shows that read-out functions cannot always be intuitively simple; on the other hand, we have shown via a ``clock TM'' that formally simple read-out functions can have unintended functionality (namely, perform the simulation themselves) that leads to a trivial notion of simulation. 

These insight have led us to formulate a version of our conjecture that defines simulation not in terms of \emph{simple}, but \emph{structure-preserving} functions: close-by configurations of the simulated TM $T$ should be represented by close-by configurations of the simulating TM $U$. While we were not able to suggest a rigorous definition of ``close-by'', we have shown that its simplest implementation leads to a notion of simulation that is not trivialized by the clock TM example. Whether this formulation of the conjecture can indeed be made rigorous in an interesting way will have to be seen in future work. 

While we were not able to settle our conjecture, we think that its study might lead to interesting insights into the nature of universal computation, regardless of whether it turns out to be true and independently of its relation to free agency. We hope that our results and constructions can motivate further interesting inquiries into the relation of computation and freedom.

\section*{Acknowledgments}
We are grateful to Andrew J.\ P.\ Garner for stimulating discussions. This research was supported by grant number FQXi-RFP-1815 from the Foundational Questions Institute and Fetzer Franklin Fund, a donor advised fund of Silicon Valley Community Foundation. MK acknowledges the support of the Vienna Doctoral School in Physics (VDSP) and the Vienna Center for Quantum Science and Technology (VCQ). This research was supported in part by Perimeter Institute for Theoretical Physics. Research at Perimeter Institute is supported by the Government of Canada through the Department of Innovation, Science and Economic Development Canada and by the Province of Ontario through the Ministry of Research, Innovation and Science. This work was also co-funded by the European Research Council (ERC) under Project No.\ 101055129. Views and opinions expressed are however those of the authors only and do not necessarily reflect those of the European Union or the European Research Council. Neither the European Union nor the granting authority can be held responsible for them.

\end{document}